\documentclass[aip,graphicx]{revtex4-1}
\usepackage{graphicx} 
\usepackage{hyperref}       %
\usepackage{amsmath}
\usepackage[version=4]{mhchem}
\usepackage{bm} 
\usepackage{amsfonts}       %
\usepackage{multirow}
\usepackage{booktabs}
\usepackage{color} 
\draft %

\begin{document}

\title{Benchmarking of quantum fidelity kernels for Gaussian process regression 
} %

\author{Xuyang Guo}
\email[ Corresponding author: ]{xyguo@chem.ubc.ca}

\affiliation{Department of Chemistry, University of British Columbia, Vancouver, B.C. V6T 1Z1, Canada}

\author{Jun Dai} 
\affiliation{Department of Chemistry, University of British Columbia, Vancouver, B.C. V6T 1Z1, Canada}

\author{Roman V. Krems}

\affiliation{Stewart Blusson Quantum Matter Institute, University of British Columbia, Vancouver, B.C. V6T 1Z4, Canada}

\date{\today}
 
\begin{abstract}
Quantum computing algorithms have been shown to produce performant quantum kernels for machine-learning classification problems. 
Here, we examine the performance of quantum kernels for regression problems of practical interest. 
For an unbiased benchmarking of quantum kernels, it is necessary to construct the most optimal functional form of the classical kernels and the most optimal quantum kernels for each given data set.  
We develop an algorithm that uses an analog of the Bayesian information criterion to optimize the sequence of quantum gates used to estimate quantum kernels for Gaussian process models. 
The algorithm increases the complexity of the quantum circuits incrementally, while improving the performance of the resulting kernels, and is shown to yield  
much higher model accuracy with fewer quantum gates than a fixed quantum circuit ansatz. We
demonstrate that quantum kernels thus obtained can be used to build accurate models of global potential energy surfaces (PES) for polyatomic molecules. The average interpolation error of the six-dimensional PES obtained with a random distribution of 2000 energy points 
is 16 cm$^{-1}$ for H$_3$O$^+$, 15 cm$^{-1}$ for H$_2$CO and 88 cm$^{-1}$ for 
HNO$_2$. We show that a compositional optimization of classical kernels for Gaussian process regression converges to the same errors. 
This indicates that quantum kernels can achieve the same, though not better, expressivity as classical kernels for regression problems. 
\end{abstract}


\maketitle %
 
\section{Introduction}
\label{sec:intro}
  
Quantum machine learning (QML) -- loosely defined as machine learning that involves a quantum computer for any part of data modeling -- is currently being researched as one of the promising applications of quantum computing. 
A popular implementation of QML encodes data into states of a quantum computer which are, after some unitary evolution, projected onto specific qubit states to estimate a kernel function for a given data set \cite{daiQuantumGaussianProcess2022,ottenQuantumMachineLearning2020,rappQuantumGaussianProcess2024, dingQuantumInspiredSupportVector2022,mafuDesignImplementationEfficient2021,
pasettoQuantumSupportVector2022,senekanePredictionSolarIrradiation2016,suzukiPredictingToxicityQuantum2020,suzukiQuantumSupportVector2023,tscharkeSemisupervisedAnomalyDetection2023,zhouQuantumKernelEstimationbased2024,
abbasQuantumEnsemblesQuantum2020,
blankQuantumClassifierTailored2020,glickCovariantQuantumKernels2022,havlicekSupervisedLearningQuantumenhanced2019,hubregtsenTrainingQuantumEmbedding2022,jagerUniversalExpressivenessVariational2023,liuRigorousRobustQuantum2021,kublerInductiveBiasQuantum2021,ningQuantumKernelLogistic2023,saeediQuantumSemisupervisedKernel2021,sancho-lorenteQuantumKernelsLearn2022,schuldQuantumMachineLearning2019,schuldQuantumModelsKernel2021,schuldSupervisedQuantumMachine2021,srikumarKernelbasedQuantumRandom2024,wangUnderstandingPowerQuantum2021,torabianCompositionalOptimizationQuantum2023}. 
Kernel functions are used in kernel machine learning methods, including support vector machines (SVM), kernel ridge regression, or Gaussian process (GP) regression.  Using a gate-based quantum computer, quantum kernels can be produced by encoding data into parameters of the quantum gates, operating with the resulting gates on qubits, and measuring the square of the amplitude of a particular state of qubits  \cite{havlicekSupervisedLearningQuantumenhanced2019}. It has been shown that this can produce quantum kernels that, when used for SVM, offer a theoretical quantum advantage for classification problems \cite{jagerUniversalExpressivenessVariational2023,liuRigorousRobustQuantum2021}. 
Many articles have explored algorithms for constructing performant quantum kernels for classification problems \cite{torabianCompositionalOptimizationQuantum2023,altares-lopezAutomaticDesignQuantum2021,chenGeneratingQuantumFeature2022,duQuantumCircuitArchitecture2022,heyraudNoisyQuantumKernel2022,incudiniStructureLearningQuantum2022,liuRepresentationLearningQuantum2022,sunkelGA4QCOGeneticAlgorithm2023,vedaieQuantumMultipleKernel2020,wangSeveralFitnessFunctions2023}. However, whether quantum kernels can be used for accurate regression models of practical interest remains largely unexplored.

Unlike the classification tasks aiming to assign a binary label (+/-) to vectors in input space, regression aims to model the dependence of a continuous output variable ($y$) on input vectors ($\bm x$). 
This can be used, for example, for building accurate models of high-dimensional potential energy surfaces for polyatomic molecules \cite{daiQuantumGaussianProcess2022}. 
 Previous work has demonstrated quantum linear regression \cite{chakrabortyQuantumRegularizedLeast2023,dateAdiabaticQuantumLinear2021a,desuAdiabaticQuantumFeature2021,duttaQuantumCircuitDesign2020,gilyenImprovedQuantuminspiredAlgorithm2022,houPartialLeastSquares2022c,kanekoLinearRegressionQuantum2021,liQuantumAlgorithmsSolving2021,liuFastQuantumAlgorithms2017,schuldPredictionLinearRegression2016b,wangQuantumAlgorithmLinear2017,zhangRealizingQuantumLinear2019,chen2021quantum,montanaro2022quantum,reddyHybridQuantumRegression2021b}, regression with quantum neural networks \cite{abelCompletelyQuantumNeural2022,beerTrainingDeepQuantum2020,diepQuantumNeuralNetworks2020,hiraiApplicationQuantumNeural2023b,kaironCOVID19OutbreakPrediction2021,killoranContinuousvariableQuantumNeural2019b,macalusoVariationalAlgorithmQuantum2020,ngoQuantumNeuralNetwork2023,paquetQuantumLeapHybridQuantum2022,qiTheoreticalErrorPerformance2023b,radDeepQuantumNeural2023,reddyHybridQuantumRegression2021b,scalaGeneralApproachDropout2023,wuExpressivityQuantumNeural2021,wuScramblingAbilityQuantum2021,verdon2019learning,diep2020nonparametric,berner2021quantum}, a quantum fuzzy regression model \cite{chenQuantumFuzzyRegression2023}, quantum-assisted regression \cite{chenFasterQuantumRidge2023b,dalalQuantumAssistedSupportVector2021,zhaoQuantumAssistedGaussian2019}, and quantum extreme learning machines \cite{Innocenti_2023,LoMonaco_2024}, where quantum algorithms are used to accelerate the training of classical regression models. It has also been shown that quantum kernels can be used as performant kernels for regression models based on support vector machines \cite{dingQuantumInspiredSupportVector2022,mafuDesignImplementationEfficient2021,pasettoQuantumSupportVector2022,senekanePredictionSolarIrradiation2016,suzukiPredictingToxicityQuantum2020,suzukiQuantumSupportVector2023,tscharkeSemisupervisedAnomalyDetection2023,zhouQuantumKernelEstimationbased2024} and Gaussian processes \cite{daiQuantumGaussianProcess2022,ottenQuantumMachineLearning2020,rappQuantumGaussianProcess2024}. 
 However, it is not clear if quantum kernels offer any advantage or can achieve similar accuracy as the most expressive classical kernels for regression problems. 
Previous work offers examples of comparison of the performance of quantum kernels with that of specific classical kernels. However, such a comparison cannot be accepted as conclusive, because the functional form of the specific classical kernels chosen for the numerical experiments may not be optimal for the data set under examination.

In order to determine whether quantum kernels can outperform classical kernels for practical applications, it is necessary to compare the performance of the most optimal quantum kernels with the most optimal classical kernels for a range of benchmark regression problems. 
To achieve this, it is necessary to align the functional form of the classical kernels and the gate sequence in the quantum circuits producing quantum kernels with the target function of each data set considered. 
As shown by Duvenaud and coworkers \cite{pmlr-v28-duvenaud13,NIPS2011_4c5bde74}, the performance of classical kernels can be systematically enhanced by increasing the complexity of the kernel function using an iterative approach combining simple kernel functions into products and linear combinations with the most optimal outcome of the previous iteration. Different kernel functions are discriminated by the value of the Bayesian information criterion (BIC) \cite{schwarzEstimatingDimensionModel1978}, serving as an easy-to-compute surrogate of marginal likelihood. BIC is an asymptotically rigorous model selection metric: if a set of models contains the target function, the probability that BIC selects the true model approaches one as the number of data points increases \cite{hastieElementsStatisticalLearning2009}. Our previous work has shown that BIC can be used as a model selection metric even in the limit of very restricted data \cite{daiInterpolationExtrapolationGlobal2020,vargas-hernandezExtrapolatingQuantumObservables2018,torabianCompositionalOptimizationQuantum2023}.

In the present work, we extend the algorithm of Duvenaud and coworkers \cite{pmlr-v28-duvenaud13} to improve the performance of quantum gate sequences used to build quantum kernels for GP regression. 
We show that the accuracy of GP regression models with quantum kernels can be systematically enhanced by increasing the number of quantum gates in the underlying quantum circuits through an iterative algorithm guided by a metric, closely related to BIC.  
This provides an algorithm to build quantum kernels for accurate GP models with the least number of quantum gates. This consequently reduces the number of gate parameters to be optimized for training ML models and reduces the overall effect of gate errors. We demonstrate this by comparing quantum circuits obtained in this work with 
the generic ansatz used previously \cite{daiQuantumGaussianProcess2022}. The present algorithm aligns quantum kernels with a given data set using the same compositional optimization strategy as the algorithm of Duvenaud and coworkers \cite{pmlr-v28-duvenaud13,NIPS2011_4c5bde74}. This allows us to improve both the classical kernels towards optimal functional form and the quantum kernels towards optimal gate sequences for a comparison of the most accurate quantum models with the most accurate classical models, for each given data set.  
The present algorithm can thus be used to expose the limitations of quantum kernels for regression problems for practical applications and to benchmark quantum kernels by classical models.

Previous studies of quantum kernels for GP regression are scarce \cite{daiQuantumGaussianProcess2022,ottenQuantumMachineLearning2020,rappQuantumGaussianProcess2024}. Gaussian processes model data with probabilistic distributions, which offer both the prediction of $y$ and the Bayesian uncertainty of the prediction. 
As such, accurate GP models with quantum kernels can be used to bridge quantum computing with a variety of new applications, ranging from efficient interpolation of molecular properties in chemical compound spaces \cite{maoEfficientInterpolationMolecular2023}, to building global PES for polyatomic molecules on a quantum computer \cite{daiQuantumGaussianProcess2022}, to developing the quantum analogs of Bayesian optimization. In the present work, we use the optimized sequence of quantum gates to build quantum kernels for GP models of PES for  H$_3$O$^+$, HNO$_2$, and H$_2$CO molecules, varying in the potential energy landscape complexity. The resulting quantum models of PES are then compared with the GP models based on classical kernels with the functional form optimized by the algorithm of Duvenaud and coworkers \cite{pmlr-v28-duvenaud13}. In addition, we compare the resulting models with neural network Gaussian process (NNGP) models, which provide an independent benchmark for both the quantum and classical model construction algorithms considered here. The results demonstrate that quantum kernels with optimized gate sequences yield GP regression models with accuracy comparable with, but not better than,  the accuracy of the most accurate classical regression models.

\clearpage

\section{Gaussian Process Regression with Classical and Quantum Kernels}
\label{sec:meth}

Gaussian process regression (GPR) is a model of the dependence of a continuous output variable $y$ on multi-dimensional input vectors $\bm x$. 
In this work, $y$ represents the potential energy of polyatomic molecules, and $\bm x$ encodes the molecular geometry. The dimensionality of $\bm x$ depends on the number of atoms in a given molecule and is equal to the number of vibrational normal modes. 
The GP model $y(\bm x)$ thus yields the global PES for the corresponding molecule.  We consider the following molecules: H$_3$O$^+$, HNO$_2$, and H$_2$CO.

GPR is a supervised learning model,  trained by a set of $N$ input-output pairs \(\mathcal{D} = \{(\bm{x}_i, {y}_i)\}_{i=1}^{N}\). The predicted potential energy at an arbitrary point $\bm{x}_*$ of input space is \cite{rasmussenGaussianProcessesMachine2005}: 

\begin{equation}
	{{y}}(\bm{x}_*) = \bm{k}^\top(\bm{x}_*) (\mathbf{K} + \sigma_n^2 \mathbf{I})^{-1} \bm{y}, 
\label{eq:gp_pred}
\end{equation} where \(\bm{k}\) is the vector of size $N$ with elements given by the kernel function values $k(\bm x_i, \bm x_*)$ and \(\mathbf{K}\) is the $N \times N$ matrix with elements given by the kernel function values $k(\bm x_i, \bm x_j)$ with $ \bm x_i, \bm x_j \in {\cal D}$. 
The prediction is parametrized by \(\sigma_n^2\) representing the variance of data noise, which is set to zero in this work since potential energy calculations are assumed to be noise-free.
 The kernel function \(k(\bm{x}, \bm{x'})\) must be positive-semidefinite and symmetric to interchange of $\bm{x}$ and $\bm{x'}$.  
 In the present work, we compare models with the kernel function  \(k(\bm{x}, \bm{x'})\) obtained using classical algorithms and from quantum states of a gate-based quantum computer. We refer to the latter as the quantum kernels. 

Once a kernel function is chosen,  the parameters of the kernel function for a GP model are estimated by maximizing the following function:
\begin{equation}
	\log \mathcal{L}({\bm \theta})=-\frac{1}{2} \boldsymbol{y}^{\top}\left(\mathbf{K}+\sigma^2 \mathbf{I}\right)^{-1} \boldsymbol{y}-\frac{1}{2} \log \left|\mathbf{K}+\sigma^2 \mathbf{I}\right|-\frac{N}{2} \log 2 \pi,
	\label{eq:lml}
\end{equation}
which is known as the type-II maximum likelihood estimation \cite{rasmussenGaussianProcessesMachine2005}. In Eq. (\ref{eq:lml}), $\bm \theta$ represents collectively all kernel function parameters and $\bm y$ is a vector of length $N$ collecting all outputs in the training set $\cal D$. 
For classical kernels, $\bm \theta$ includes all free parameters of the mathematical expression for a given kernel function. 
For quantum kernels, $\bm \theta$ represents parameters of quantum gates, as described in detail below. 
In order to benchmark quantum kernels, it is necessary to consider classical kernels with varying function complexity in order to build the most optimal classical GP models for a given data set. Here, we employ two algorithms for aligning classical kernels of GP models with given data: the algorithm of 
Duvenaud and coworkers \cite{pmlr-v28-duvenaud13}, yielding composite kernels, and neural network Gaussian processes \cite{daiNeuralNetworkGaussian2023}. 
 
 \subsection{Classical composite kernels}
 \label{sec:cc}
Most traditional applications of kernel methods use simple mathematical expressions for kernel functions, such as the radial basis function (RBF): 
 \begin{equation}
	k\left(\boldsymbol{x}, \boldsymbol{x}^{\prime}\right)=\exp \left(-\theta\left\|\boldsymbol{x}-\boldsymbol{x}^{\prime}\right\|^2\right), 
\label{eq:rbf}
\end{equation} where $\theta$ is the kernel function parameter that can be tuned to train the corresponding model. The RBF kernel is known to be universal\cite{wangRBFKernelBased2004} and represents one of the most commonly used kernel functions in kernel ML. 
We use the RBF kernel as the benchmark example of simple classical kernels.

\begin{figure*}
	\centering
	\includegraphics[width=\textwidth]{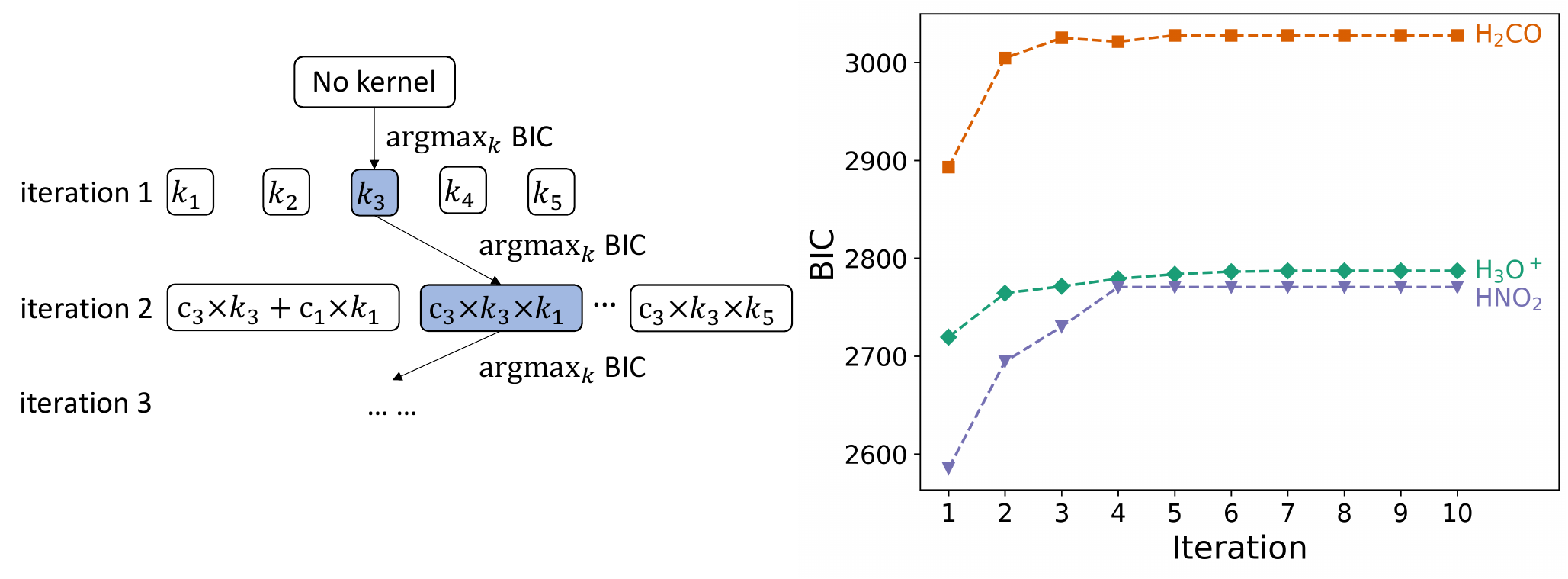}
	\caption{\label{fig:composite_classical} Left: Schematic illustration of the iterative process of composite classical kernel construction. At each iteration, the kernel with the largest value of the Bayesian information criterion (BIC) is chosen as the optimal kernel. The functions $k_i$ with $i \in [1,5]$ correspond to the kernel functions given in Eqs. (\ref{eq:rbf})-(\ref{eq:matern}). Right: Dependence of the BIC value for GP models of six-dimensional PES for the molecules indicated in the curve labels on the kernel complexity level. Each model is trained with 500 values of the potential energy randomly sampled from the configuration space of the corresponding molecule. 
	}
\end{figure*}

In addition, we build classical models with composite kernel functions optimized to maximize the Bayesian information criterion as proposed by Duvenaud and coworkers \cite{NIPS2011_4c5bde74,pmlr-v28-duvenaud13}. 
This approach can be used to increase the kernel function complexity in order to improve the resulting GP models. The algorithm for building composite kernels  
was described in Refs. \onlinecite{daiInterpolationExtrapolationGlobal2020,vargas-hernandezExtrapolatingQuantumObservables2018,pmlr-v28-duvenaud13,NIPS2011_4c5bde74} and is illustrated in Figure \ref{fig:composite_classical}. 
For a given dataset, the kernel selection process begins with a set of simple (base) kernel functions, including Eq. (\ref{eq:rbf}) and the following functions:
\begin{equation}
	k\left(\boldsymbol{x}, \boldsymbol{x}^{\prime}\right)=\boldsymbol{x}^{\mathrm{T}} \boldsymbol{x}^{\prime}
	\label{eq:dot}
\end{equation}
\begin{equation}
	k\left(\boldsymbol{x}, \boldsymbol{x}^{\prime}\right)=\left(1+\frac{d\left(\boldsymbol{x}, \boldsymbol{x}^{\prime}\right)^2}{2 \alpha l^2}\right)^{-\alpha}
	\label{eq:rq}
	\end{equation}
\begin{equation}
	k\left(\boldsymbol{x}, \boldsymbol{x}^{\prime}\right)=\exp \left(-\frac{2 \sin ^2\left(\pi d\left(\boldsymbol{x}, \boldsymbol{x}^{\prime}\right) / p\right)}{l^2}\right)
	\label{eq:exp}
\end{equation}
\begin{equation}
	k\left(\boldsymbol{x}, \boldsymbol{x}^{\prime}\right)=\frac{2^{1-v}}{\Gamma(v)}\left(\sqrt{2 v} r\left(\boldsymbol{x}, \boldsymbol{x}^{\prime}\right)\right)^v \mathcal{K}_v\left(\sqrt{2 v} r\left(\boldsymbol{x}, \boldsymbol{x}^{\prime}\right)\right)
	\label{eq:matern}
\end{equation}
where $d\left(\boldsymbol{x}, \boldsymbol{x}^{\prime}\right)$ is the Euclidean distance between $\boldsymbol{x}$ and $\boldsymbol{x}^{\prime}$, $\alpha$ is a positive scale mixture parameter, $l$ is a positive length-scale parameter, $p$ is a positive periodicity parameter, $\Gamma$ is the gamma function, $v$ is a non-negative half-integer, and $\mathcal{K}_v$ is the modified Bessel function. These kernel functions are
represented in Figure \ref{fig:composite_classical} as $k_i$ with $i \in [1,5]$. The kernel functions are discriminated by the value of 
Bayesian information criterion (BIC) defined as \ref{eq:bic}:
\begin{equation}
	\mathrm{BIC}=\log \mathcal{L}(\hat {\bm \theta})-\frac{1}{2} \mathcal{M} \log N,
	\label{eq:bic}
\end{equation}
where 
$\log \mathcal{L}(\hat {\bm \theta})$ is the maximum value of the function in Eq. (\ref{eq:lml}) and 
$\mathcal{M}$ is the number of parameters in the kernel function. 
In the following iteration, the base kernel with the largest value of BIC ($k_{\rm opt}$) is combined with each base kernel as linear combinations $c_i k_{\text{opt}}+c_j k_j$ and products $c_ik_{\text{opt}}\times k_j$, resulting in 10 new kernel functions. These kernel functions are optimized by varying both $\bm \theta$ and the coefficients $c_i$ to maximize Eq. (\ref{eq:lml}).  The kernel with the largest value of BIC is then selected as the optimal kernel and the process is iterated. This iterative process continues until the BIC value of the optimal kernel converges, as shown in the right panel of Figure~\ref{fig:composite_classical}, for several examples of GP models of PES for the molecules labeling the curves.

\subsection{Neural network Gaussian processes}

To provide an independent benchmark for quantum kernels, we also consider neural network Gaussian process (NNGP) models \cite{daiNeuralNetworkGaussian2023,garriga-alonsoDeepConvolutionalNetworks2019,leeDeepNeuralNetworks2018,matthewsGaussianProcessBehaviour2018,novakBayesianDeepConvolutional2020,sohl-dicksteinInfiniteWidthLimit2020,neuraltangents2020}.
NNGP models exploit the connection between GPs and neural networks (NN). The output of a single-layer, fully connected feed-forward neural network is given by 
\begin{equation}
y(\bm{x})=b+\sum_{j=1}^{N} W_{j} y_{j}(\bm{x}), 
\label{NN_1}
\end{equation}
with 
\begin{equation}
\quad y_{j}(\bm{x})=\phi\left( b_{j}^{0}+\sum_{i=1}^{p} W_{i j}^{0} x_i\right),
\end{equation}
where $x_i$ is the $i$th component of the $p$-dimensional vector $\bm x$, $W_{i j}$ and $ b_0$ are the weight and bias parameters for the layer, $\phi(y)$ is a non-linear activation function,  and $N$ is the number of nodes.

If the weights and bias parameters are chosen to be random variables,  the central limit theorem ensures that, 
in the limit  of infinite width $N \to \infty$ with priors $\mathcal{N}(\mu_w, {\sigma_{w}}^{2})$ and $\mathcal{N}(\mu_b, {\sigma_{b}}^{2})$,  such NN becomes a Gaussian process, 
\begin{eqnarray}
y(\bm x) \sim \mathcal{GP}(\mu,k)
\end{eqnarray} 
with mean $\mu$ and covariance $k$ functions. Given the freedom of the priors, one can choose $\mu_w = \mu_b = 0$, yielding $\mu=0$ and 
\begin{equation}
 k(\bm{x},\bm{x'}) = \mathbb{E}\left[y\left(\bm{x}\right) y\left(\bm{x'}\right)\right] =\sigma_{b}^{2}+\sum_{j} \sigma_{w}^{2} ~\mathbb{E} \left[y_j\left(\bm{x}\right) y_j\left(\bm{x'}\right)\right].
\label{K_1}
\end{equation}

It was recently shown that this can be extended to NN with multiple hidden layers \cite{leeDeepNeuralNetworks2018,matthewsGaussianProcessBehaviour2018,garriga-alonsoDeepConvolutionalNetworks2019,sohl-dicksteinInfiniteWidthLimit2020,neuraltangents2020}. 
The variables $y_j$ and $N$ must now be labeled by the layer index $l \in [1, L]$. Given that each of $y_j^{l-1}$, used as inputs into node $i$ of layer $l$, is an independent GP, as $N_l \to \infty$, $y_i^l$ is also a Gaussian process $\mathcal{GP}(0,k^l)$ with the covariance function

\begin{equation}
 k^l(\bm{x},\bm{x'}) = \mathbb{E}\left[y_i^l\left(\bm{x}\right) y_i^l\left(\bm{x'}\right)\right] ={\sigma^{(l)}_{b}}^{2}+ {\sigma^{(l)}_{w}}^{2}\mathbb{E}_{y_i^{l-1} \sim \mathcal{GP}(0,K^{l-1})} \left[\phi(y^{l-1}_i\left(\bm{x}\right)) \phi(y^{l-1}_i\left(\bm{x'}\right))\right].
\label{K_l}
\end{equation}
Any two $y_i^l$ and $y_j^l$ in the same layer $l$ of this NN are independent.  The recursive form of $k^l$ depends on the activation function $\phi$ and the parameters $\sigma^{(l)}_{w}$ and $\sigma^{(l)}_{b}$. Here, we use $ \phi(y) = \mathrm{erf}(y) = \frac{2}{\sqrt{\pi}} \int_{0}^{y} e^{-t^{2}} d t$ and treat $\sigma^{(l)}_{w}$ and $\sigma^{(l)}_{b}$ for each layer $l$ as independent trainable parameters. 
The parameters $\sigma^{(l)}_{w}$ and $\sigma^{(l)}_{b}$ for each layer are optimized by type-II maximum likelihood estimation \cite{rasmussenGaussianProcessesMachine2005} using Bayesian optimization. The number of layers $L$ of NNGP is increased until the magnitude of the function in Eq. (\ref{eq:lml}) converges.

\subsection{Quantum kernels}
\label{sec:qk}
Quantum kernels considered here are given by \begin{equation}
	k\left(\boldsymbol{x}, \boldsymbol{x}^{\prime}\right)=\left|\left\langle\psi_0\left|\mathcal{U}^{\dagger}\left(\boldsymbol{x}^{\prime}\right) \mathcal{U}(\boldsymbol{x})\right| \psi_0\right\rangle\right|^2, 
\label{eq:qk}
\end{equation} where  $\mathcal{U}\left(\boldsymbol{x}\right)$ is a unitary transformation given by a sequence of one- and two-qubit gates acting on a quantum state of $m$ qubits, and 
$|\psi_0\rangle = |0\rangle^{\otimes m}$. Note that  $\mathcal{U}\left(\boldsymbol{x}\right)$ is parametrized by $\bm x$ so the quantum sates \(\mathcal{U}\left(\boldsymbol{x}^{\prime}\right)| \psi_0\rangle \) and \(\mathcal{U}(\boldsymbol{x})| \psi_0\rangle\) depend on the input vectors $\bm x'$ and $\bm x$. 
We encode each dimension of the molecular geometry vector $\bm{x}$ into a separate qubit.
The number of qubits required to build quantum states  \(\mathcal{U}(\boldsymbol{x})| \psi_0\rangle\) is therefore equal to the dimensionality of the configuration space of the molecule.  
 A sequence of gate operations $\mathcal{U}^{\dagger}\left(\boldsymbol{x}^{\prime}\right) \mathcal{U}(\boldsymbol{x})$ is then applied to the initial state $|0\rangle^{\otimes m}$. The kernel function $k\left(\boldsymbol{x}, \boldsymbol{x}^{\prime}\right)$ is obtained by the projection of the output state $\mathcal{U}^{\dagger}\left(\boldsymbol{x}^{\prime}\right) \mathcal{U}(\boldsymbol{x})|0\rangle^{\otimes m}$ onto the original state $|0\rangle^{\otimes m}$.

\begin{figure*}
	\centering
	\includegraphics[width=\textwidth]{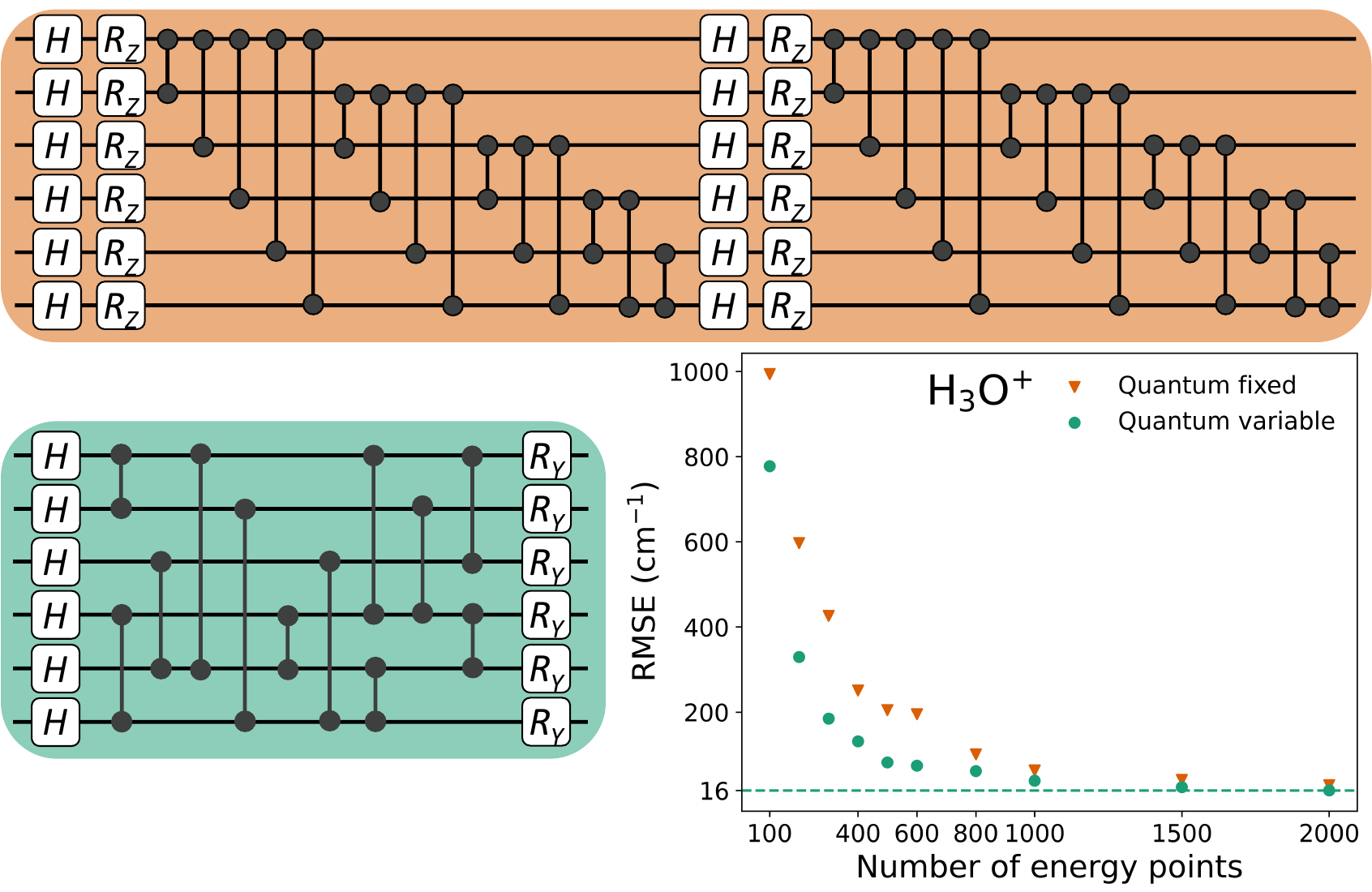}
	\caption{\label{fig:depth_perf} 
	Upper panel: Quantum circuit $\cal U$ for the fixed ansatz used for GP regression in Ref. \onlinecite{daiQuantumGaussianProcess2022} that yields orange triangles in the lower right panel.
	Lower left: Quantum circuit $\cal U$ for the optimal quantum kernel obtained by the compositional optimization algorithm described in the text. The ansatz is identified with 500 training points randomly sampled from the entire energy range and yields blue circles in the lower right panel. 
	Lower right: RMSE of the GP model of PES for H$_3$O$^+$ with the variable ansatz (blue circles) and fixed ansatz (orange triangles) for the quantum kernels as a function of the number of training points. 
	 $H$ denotes the Hadamard gate, $R_Z$ -- the $R_Z$ gate and $R_Y$ -- the $R_Y$ gate. All entangling gates are $R_{ZZ}$. 
	}
\end{figure*}

The performance of the quantum kernels in ML models is determined by the specific sequence of gate operations in  $\mathcal{U}$.  
Previous work on GP regression with quantum kernels \cite{daiQuantumGaussianProcess2022} used a fixed ansatz $\mathcal{U}$ 
given by 
\begin{eqnarray}
\mathcal{U} = UH^{\otimes m}UH^{\otimes m},
\label{old-ansatz}
\end{eqnarray} 
where $H$ is the single-qubit Hadamard operator,  
\begin{equation}
	U=\exp \left[-i\left(\sum_i^m \sigma_{Z, i} \phi_i+\sum_{i, j>i}^m \sigma_{Z, i} \sigma_{Z, j} \phi_{i j} \right)\right],
	\end{equation} $\sigma_{Z, i}$ is the Pauli Z-matrix operating on qubit $i$, and $\phi_i$ and $\phi_{i j}$ are the quantum gate parameters. 
The quantum circuit for the quantum kernel based on this ansatz is shown in the upper panel of Figure~\ref{fig:depth_perf} for the example of a six-dimensional problem (a molecule with four atoms). 
The quantum circuit in Figure~\ref{fig:depth_perf} (top panel) is depicted as a sequence of Hadamard gates, $R_Z$ gates defined as
\begin{eqnarray}
R_{Z} = \exp(-i \frac{\phi}{2}\sigma_Z),
\end{eqnarray}
and entangling $R_{ZZ}$ gates defined as
\begin{eqnarray}
R_{ZZ} = \exp(-i \frac{\phi}{2}\sigma_Z \otimes \sigma_Z),
\end{eqnarray}
where the parameters $\phi$ depend on the qubit index for $R_Z$ and two qubit indices for $R_{ZZ}$. 
The ansatz shown in Eq.~\ref{old-ansatz} was borrowed from previous work on QML for classification \cite{havlicekSupervisedLearningQuantumenhanced2019} and is considered to be general, as it provides entanglement between all pairs of qubits. 
Our previous work showed \cite{daiQuantumGaussianProcess2022} that this ansatz yields quantum kernels for GP regression of a six-dimensional PES with accuracy comparable to that attainable by GP regression models with the RBF kernel. 

The present work aims to develop an algorithm for optimizing the gate sequences in $\mathcal{U}$ to adapt the quantum kernels to a given data set. This can be viewed as a search problem in the space of all possible gate permutations. The complexity of the search increases exponentially with the number of gates in the quantum circuits. To accelerate the search, we use the strategy inspired by the algorithm of Duvenaud and coworkers \cite{NIPS2011_4c5bde74,pmlr-v28-duvenaud13,vargas-hernandezExtrapolatingQuantumObservables2018,daiInterpolationExtrapolationGlobal2020}. 
We previously observed that a similar algorithm can be used to enhance the performance of quantum kernels for classification problems with SVM \cite{torabianCompositionalOptimizationQuantum2023}.

\begin{figure*}
	\centering
	\includegraphics[width=0.8\textwidth]{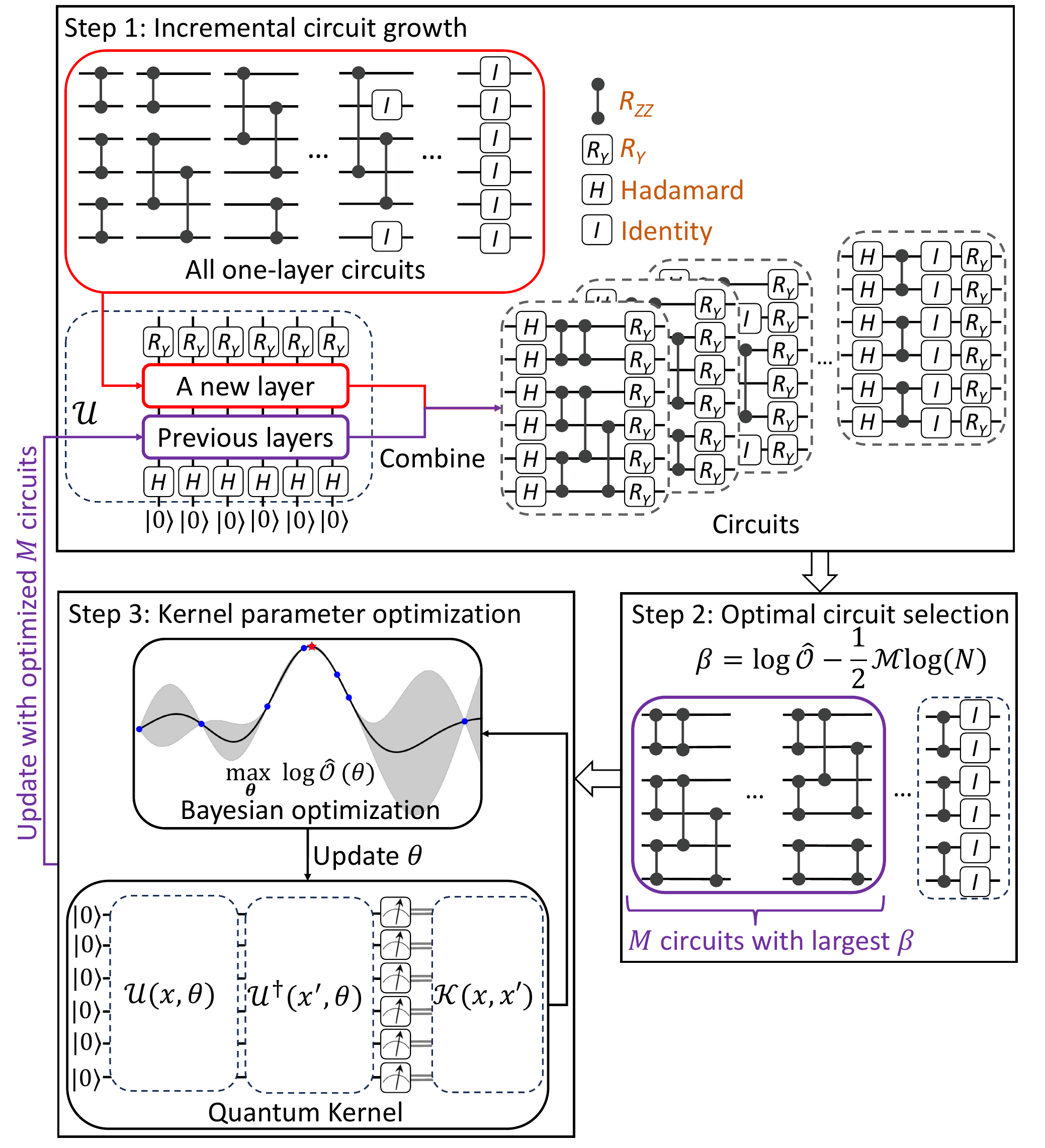}
	\caption{\label{fig:comp_search} Schematic diagram of the compositional search of quantum circuits used in the present work. The algorithm increases the number of layers in quantum kernels iteratively and includes three steps: (1) $M$ best circuits from the previous iteration are appended with the pool of one-layer circuits to generate a pool of circuits with a greater depth; (2)  
	$M$ quantum circuits with the largest values of $\beta$ as defined by Eq. {\ref{eq:beta}} are retained for the next iteration; (3) The trainable parameters of each circuit are optimized by Bayesian optimization to maximize Eq. (\ref{eq:o_lml}) on a training set with randomly sampled \textit{ab inito} points from the entire energy range of the target PES.
	}
\end{figure*}

 The algorithm is based on the following three gates as the building blocks: the Hadamard gate $H$, the $R_{ZZ}$ gate, and the $R_Y$ gate defined as
\begin{eqnarray}
R_{Y} = \exp(-i \frac{\phi}{2}\sigma_Y)
\end{eqnarray}
where $\sigma_Y$ is the Pauli $Y$-matrix and the parameter $\phi$ depends on the index of the qubit involved in the transformation.
The unitary transformation $\mathcal{U}$ is structured, to begin with, a layer of Hadamard gates $H^{\otimes m}$, followed by parameterized entanglement layers $U_e$ including a sequence of \(R_{ZZ}\) gates and identity gates, and to conclude with a parameterized layer of \(R_Y\)  gates:
\begin{equation}
	\mathcal{U} =  R_Y^{\otimes m}U_eH^{\otimes m}.  
	\label{optimized-u}
\end{equation}
The sequence of gates in $U_e$ is determined by the compositional search algorithm described below and illustrated in Fig.~\ref{fig:comp_search}.

The input vectors $\boldsymbol{x}$ are encoded into the quantum gates as follows:
\begin{equation}
	\begin{array}{cc}
	\phi_i=x_i / \theta_i~~\forall~R_Y, \\
	\phi_{i j}=\exp \left(-\left(x_i-x_j\right)^2 / \theta_{i j}\right)~~\forall~R_{ZZ},
	\end{array}
\end{equation}where  $x_i$ and $x_j$ are the $i$th and $j$th components of vector $\bm x$, and $\theta_i$ and $\theta_{i j}$ are the trainable parameters of a quantum kernel. The parameters $\theta_{i}$ for single qubit rotation gates are independent of each other, while all parameters $\theta_{i j}$ of the two-qubit gates $R_{ZZ}$ gates are set equal to a single variable $\Theta$. All of the quantum kernel parameters collectively denoted by $\bm{\theta} = [\theta_{i=1, \ldots, 6}, \Theta]^{\top}$ are determined by the type-II maximum likelihood estimation\cite{rasmussenGaussianProcessesMachine2005}.  

As was previously observed \cite{daiQuantumGaussianProcess2022}, the likelihood functions $\cal L$ can be very small for a large number of quantum circuit parameters $\bm \theta$. 
As a result, the function defined by Eq. (\ref{eq:lml}) exhibits a large number of divergences, which complicate the numerical optimization. Therefore, as in Ref. \onlinecite{daiQuantumGaussianProcess2022}, we optimize 
\begin{equation}
	\mathcal{O}(\boldsymbol{\theta})=\log [\mathcal{L}(\boldsymbol{\theta})+d],
\label{eq:o_lml}
\end{equation} 
instead of the function in Eq. (\ref{eq:lml}) directly in order to train quantum kernels. We set $d=1$. 
All quantum circuits in the present work are simulated by Statevector in the IBM Qiskit package\cite{treinishQiskitQiskitmetapackageQiskit2023} and hence there is no error or noise in quantum operations. Following 
Ref.~\onlinecite{daiQuantumGaussianProcess2022},
we use Bayesian optimization to maximize the function in Eq. (\ref{eq:o_lml}) to obtain optimized parameters for quantum kernels. For more details on training quantum kernels, see Ref.~\onlinecite{daiQuantumGaussianProcess2022}. 

\begin{figure*}
	\centering
	\includegraphics[width=0.46\textwidth]{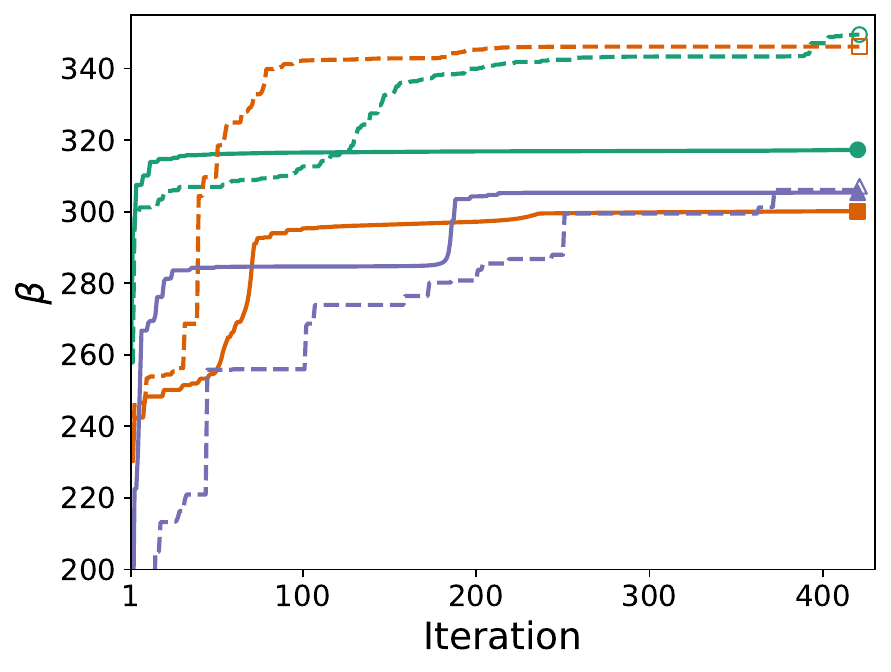}
	\includegraphics[width=0.46\textwidth]{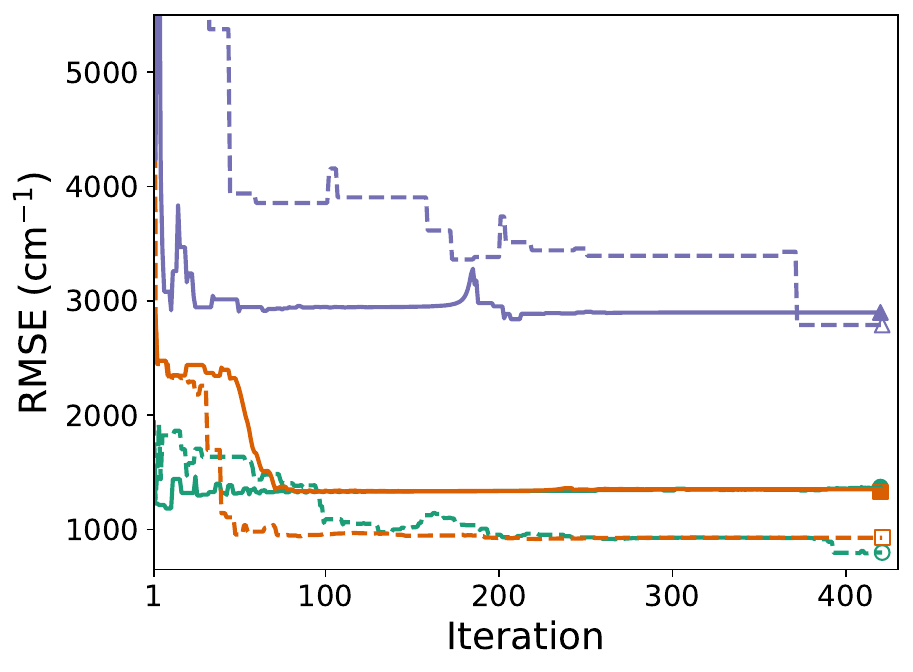}
	\caption{\label{fig:bo_vs_adam} Optimization convergence of the quantum kernel seletion metric $\beta$ defined by Eq.~(\ref{eq:beta}) (left panel) and RMSE (\ref{RMSE_equation}) (right panel) for GP models of 6D PES for H$_3$O$^+$ (green curves converging to circles), HNO$_2$ (orange curves converging triangles), and H$_2$CO (purple curves converging to squares): solid curves and full symbols -- the Adam gradient descent algorithms; broken curves and open symbols -- Bayesian optimization.
	 Each model is trained with 100 values of the potential energy randomly sampled from the configuration space of the corresponding molecule. }
\end{figure*}

	To evaluate the performance of Bayesian optimization, we repeated optimization of GP models with the quantum kernel based on the fixed ansatz in Eq.~(\ref{old-ansatz}) with the Adam gradient descent algorithm~\cite{kingma2017adammethodstochasticoptimization}. The comparison of Bayesian optimization and gradient-based optimization for training QGP models is illustrated in Fig.~\ref{fig:bo_vs_adam}. To compute the gradients of $\beta$ in Eq.~(\ref{eq:beta}) with respect to the kernel parameters, we use back-propagation by automatic differentiation as implemented in the Pennylane package~\cite{bergholm2022pennylaneautomaticdifferentiationhybrid}. The results show that the gradient descent optimization rapidly converges to local exterma for H$_3$O$^+$ (green curves) and H$_2$CO (purple curves), while Bayesian optimization leads to larger values of $\beta$ and higher interpolation accuracy of the QGP models. For a more complex PES of HNO$_2$, the gradient descent optimization converges faster and to the same limit as Bayesian optimization. Both optimizers reach similar values of $\beta$ and interpolation accuracy with 400 iterations. We note that the back-propagation by automatic differentiation is more costly than Bayesian optimization as it requires additional computation resources to store tensors and calculate the Jacobian matrix for each step in the calculation of $\beta$ \cite{bergholm2022pennylaneautomaticdifferentiationhybrid}.

\section{Compositional optimization of quantum gate sequences}
\label{sec:comp_search}

To build an optimal unitary transformation $U_e$ in Eq.~(\ref{optimized-u}), we propose the strategy illustrated 
in Fig.~\ref{fig:comp_search}.  This algorithm increases iteratively the complexity of $U_e$ and hence improves the expressivity of the corresponding quantum kernel.  
As shown in Fig.~\ref{fig:comp_search}, the algorithm is controlled by 
 hyperparameter $M$, which controls the number of quantum circuits retained in each iteration. As $M\rightarrow \infty$, the algorithm approaches full search in the space of gate permutations. 

 Duvenaud and coworkers \cite{NIPS2011_4c5bde74,pmlr-v28-duvenaud13,vargas-hernandezExtrapolatingQuantumObservables2018,daiInterpolationExtrapolationGlobal2020} used 
 the BIC defined in Eq.~(\ref{eq:bic}) as the model selection metric.
As mentioned above, however, $\log {\cal L}$ is unsuitable for optimizing quantum circuit parameters. 
For numerical stability, we use the following function as the quantum circuit selection metric:
 \begin{equation}
	\beta =\log \mathcal{\mathcal{\hat O}}-\frac{1}{2} \mathcal{M} \log N,
\label{eq:beta}
\end{equation} 
where $\mathcal{\hat O}$ is the maximum value of the function given by Eq.~\eqref{eq:o_lml}.
As $\log {\cal O}$ and $\log {\cal L}$ are maximized by the same set of parameters, $\beta$ is analogous to BIC.

As illustrated in Figure~\ref{fig:comp_search}, the algorithm is initialized by a pool of single-layer circuits, which contain all possible combinations of $R_{ZZ}$ and identity gates for a given number of qubits -- a total of $J$ circuits. To reduce the computational cost, we require that each qubit can only be operated by one gate in each layer. 
Each iteration of the algorithm includes three steps. 
First, the algorithm combines $M$ circuits with the largest value of $\beta$ in Eq.~(\ref{eq:beta}) with all possible single-layer circuits to generate $M \times J$ new circuits. As shown by Eq.~(\ref{optimized-u}), each new circuit is sandwiched between a layer of Hadamard gates and a layer of $R_Y$ gates. Second, each wrapped circuit is parameterized by the optimized parameters from the $M$ selected circuits. The quantum kernel selection metric $\beta$ is computed for each circuit and $M$ circuits with the largest $\beta$ are selected for the next iteration. Third, using the previously determined parameters as initial guesses, the kernel parameters of each selected circuit are optimized by maximizing $\log \mathcal{\hat O}$ using Bayesian optimization. The iterative process continues until the largest value of $\beta$ converges. After the convergence, all kernel parameters of the circuit with the largest $\beta$ are re-optimized to obtain the optimal parameterized circuit.

A similar algorithm was previously used to improve the performance of quantum kernels for classification with SVM \cite{torabianCompositionalOptimizationQuantum2023}. 
However, there are some important differences between classification with SVM and GP regression, which leads to the following important differences in the compositional search of quantum circuits.  
For SVM models, the architecture of quantum circuits can be optimized separately from the parameters of the quantum circuits. This simplifies the compositional search of quantum circuits to a great extent. 
For regression problems, on the other hand, the parameters of the quantum kernels must be optimized at every step of the circuit selection algorithm (Step 2 in Figure~\ref{fig:comp_search}). 
To reduce the complexity of the problem, we restrict the choice of gates for $U_e$ to $R_{ZZ}$ and identity operators and allow only one gate per qubit per layer. 
In addition, the outputs of SVM must be converted to probabilistic predictions to allow the computation of BIC on a validation set. In the context of GP regression, no validation set is necessary as BIC, or equivalently the value of $\beta$ as defined by Eq. (\ref{eq:beta}), is a byproduct of training 
the GP model by type-II maximum likelihood estimation.

\clearpage
\newpage

\section{Numerical results}

 To illustrate the algorithm for the compositional optimization of quantum kernels described in the previous section and benchmark the performance of the quantum kernels, we consider three six-dimensional regression problems yielding global PES for 
 the following molecules: H$_3$O$^+$, HNO$_2$, and H$_2$CO. 
The molecular geometry is described by six-dimensional vectors $\bm x$. The components of $\bm x$ are defined as $x_i = \exp(-{r_{i}}/{a})$, where $r_{i}$ is one of six atom-atom distances within a molecule and the parameter
$a$ is fixed to 2.5 \cite{yuInitioPotentialStep2016} for H$_3$O$^+$ and 1.0 for all other molecules. 

 The GP models are trained and tested by the potential energy of the molecules computed in Refs. \onlinecite{yuInitioPotentialStep2016,schmitzMachineLearningPotential2019}.
The potential energy data include 31124, 77272, 77284 \textit{ab initio} points spanning the energy range [0, 21000] cm$^{-1}$ for H$_3$O$^{+}$, [0, 44000] cm$^{-1}$ for H$_2$CO, and [0, 36000] cm$^{-1}$ for HNO$_2$, respectively. GP models are trained with $N$ \textit{ab initio} points that are randomly sampled from a specific energy interval.
 
 The compositional optimization of gate sequences for quantum kernels is performed with a small number of 
potential energy points (300, 400, or 500, as specified in the corresponding figure captions) randomly sampled from the configuration space of the corresponding molecule. The model accuracy is quantified by the root-mean-squared error, 
\begin{equation}\label{RMSE_equation}
{\rm RMSE} = \sqrt{\frac{\Sigma_{i=1}^{N_{\rm test}}(y_{i}-\hat{y_{i}})^2}{N_{\rm test}}}
\end{equation}
where $\hat y_i$ denotes model predictions and $y_i$ are the corresponding potential energy points from Refs.  \onlinecite{yuInitioPotentialStep2016,schmitzMachineLearningPotential2019}. The sum in Eq. (\ref{RMSE_equation}) extends over all {\it ab initio} points that are not used for training the models.

\subsection{Variable vs fixed ansatz}

Figure \ref{fig:depth_perf} compares the quantum circuits of the fixed ansatz in Eq.~(\ref{old-ansatz}) with the most optimal outcome of the compositional optimization for the model of PES for the molecule H$_3$O$^+$. 
The lower left panel of Figure \ref{fig:depth_perf}  shows the architecture of quantum kernel $\cal U$ produced by the compositional optimization algorithm and the lower right panel compares the RMSE of the GP model predictions with the quantum kernels from two $\cal U$ circuits shown. 
Figure \ref{fig:depth_perf}  demonstrates that the compositional optimization produces quantum kernels that both require fewer gates and yield more accurate regression models. 

\begin{figure*}
	\centering
	\includegraphics[width=\textwidth]{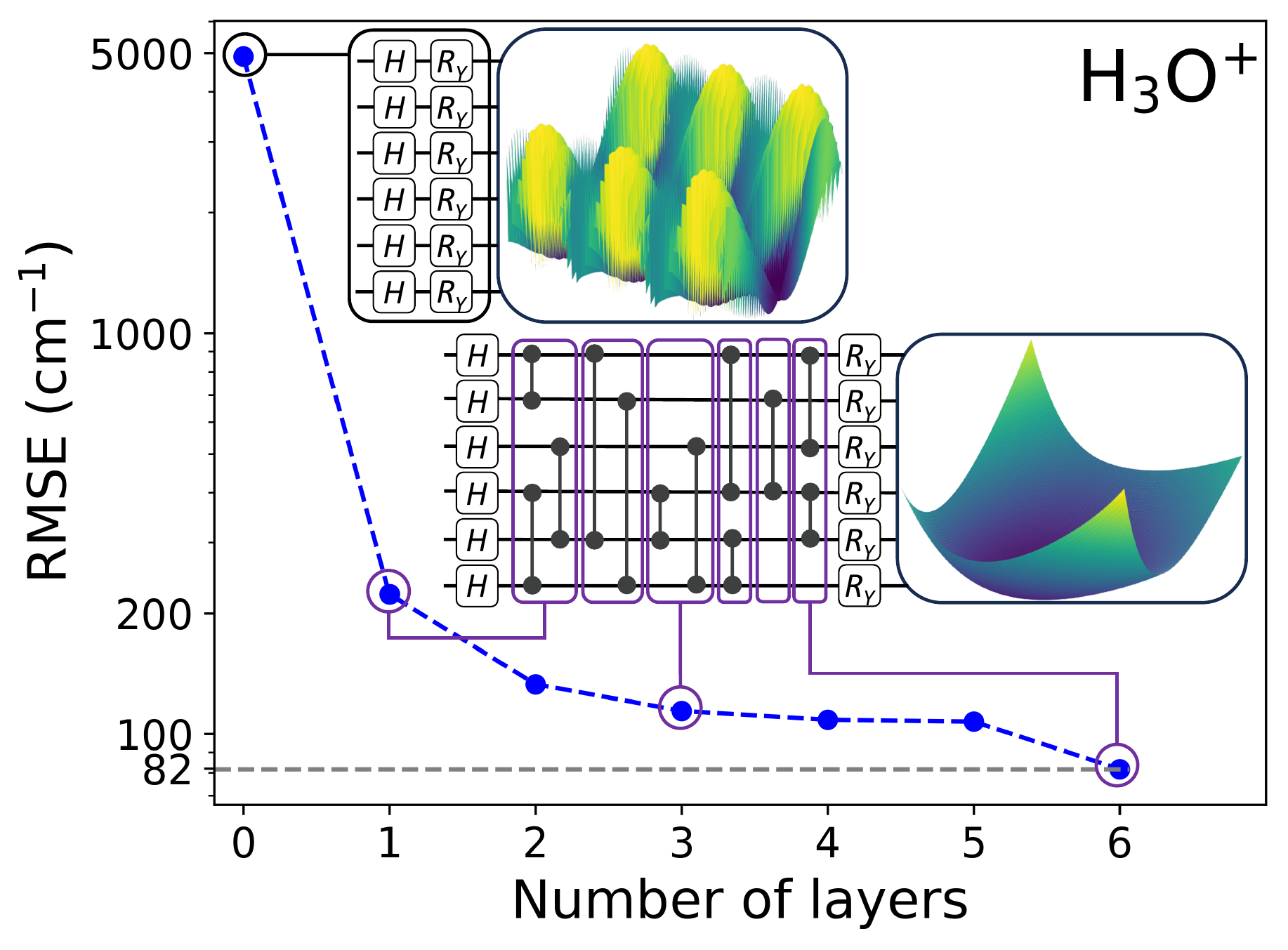}
	\caption{\label{fig:model_building}
	Improvement of the regression model of PES for H$_3$O$^{+}$ based on quantum kernels with optimized architecture with the increasing complexity of the quantum circuits.  The six-dimensional models of PES are trained by 500 potential energy points. 
	The insets show the potential energy of the molecule as a function of the shortest H-O distance and the distance between two oxygen atoms.}
\end{figure*}

Figure \ref{fig:model_building} illustrates the improvement of the regression model of the global PES for H$_3$O$^+$ based on quantum kernels with optimized architecture with increasing complexity of the quantum circuits. It can be observed that the PES model with $\cal U$ including only the Hadamard 
and $R_Y$ gates is completely unphysical, while the model based on $\cal U$ with 6 layers in $U_e$ produces a smooth PES with a small RMSE. 
It can be seen that the circuit optimization rapidly reaches convergence, indicating that physical PES models can be obtained with shallow quantum circuits. 
This rapid convergence to an optimized circuit architecture illustrates the efficiency of the kernel optimization algorithm, highlighting its ability to identify important entanglement patterns and circuit structures that significantly enhance the interpolation accuracy of the QGP models. 
The lowest value of the RMSE in Figure  \ref{fig:model_building} is 82 cm$^{-1}$. Note that the models of the global 6D PES illustrated in Figure \ref{fig:model_building} are obtained with 500 potential energy points. The error of ML models scales with the number of training points $N$ as $\propto 1/\sqrt{N}$. 

\subsection{Comparison of quantum kernels with classical kernels}

\begin{figure*}
	\centering
	\includegraphics[width=\textwidth]{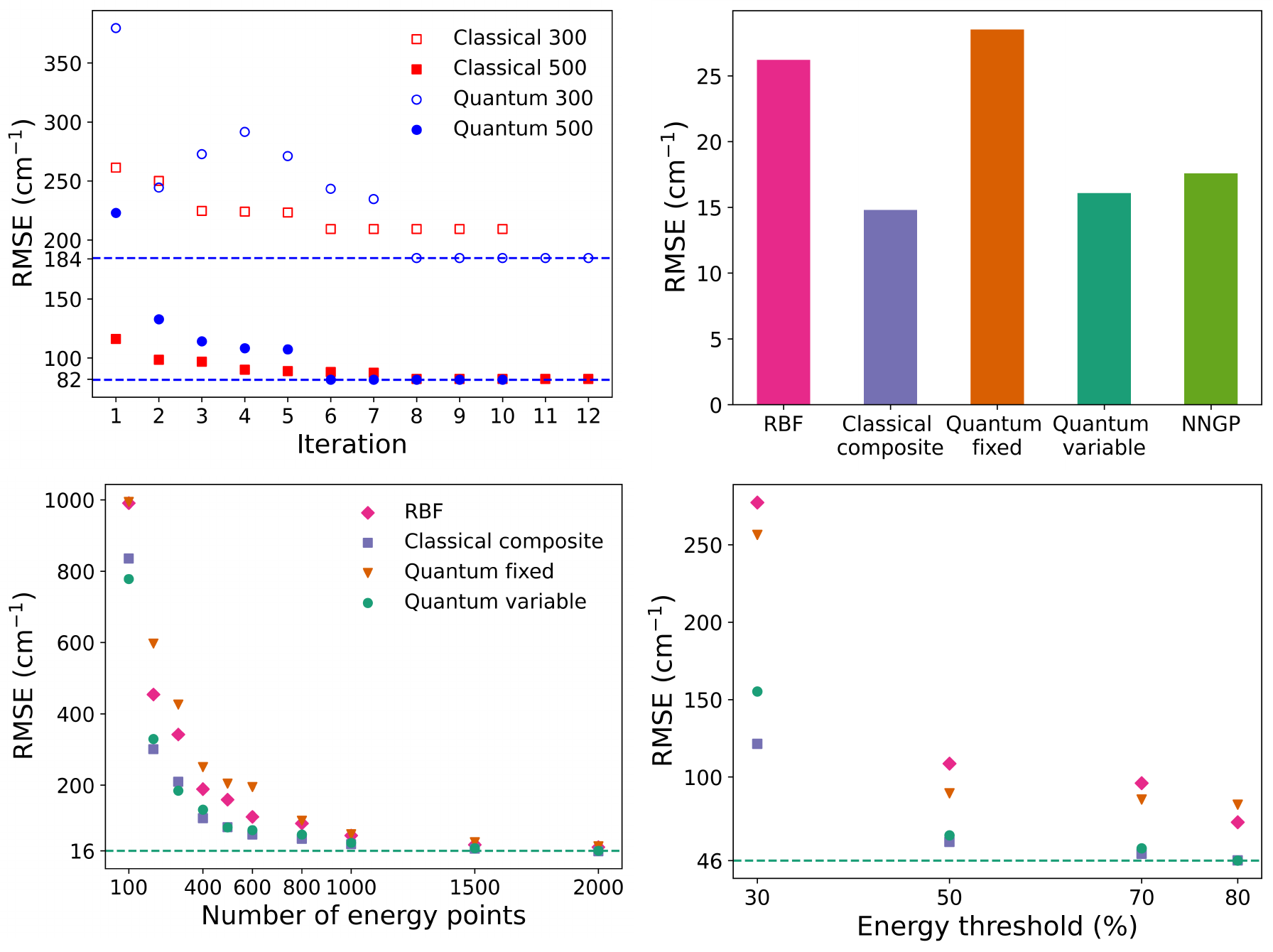}
	\caption{\label{molecule1} Upper left: Convergence of RMSE for the interpolation models yielding global PES for H$_3$O$^+$ with increasing complexity of composite classical and quantum kernels with adaptive ansatz. The functional form of the classical kernels and the architecture of the quantum kernels are optimized with 300 or 500 energy points, as indicated. The identical distributions of energy points are used for optimizing the kernels for both classical and quantum models and for training the resulting GP models. 
	Lower left: RMSE vs the number of training points for GP interpolation models of PES for H$_3$O$^+$. The structure of the kernels (for composite classical and quantum variable) is determined with the number of energy points indicated on the $x$-axis. The samples of these energy points are selected to be identical for each kernel, but random for each number of data points considered. 
	Upper right: Comparison of the lowest RMSE for the GP models of global PES for H$_3$O$^+$ 
	constructed with five different kernels, as specified in the figure. The RMSE shown corresponds to the lowest RMSE in the lower left panel.  Lower right: RMSE of GP models for extrapolation of PES in the energy domain constructed with the same kernels as discussed in the text. 
	}
\end{figure*}

\begin{figure*}
	\centering
	\includegraphics[width=\textwidth]{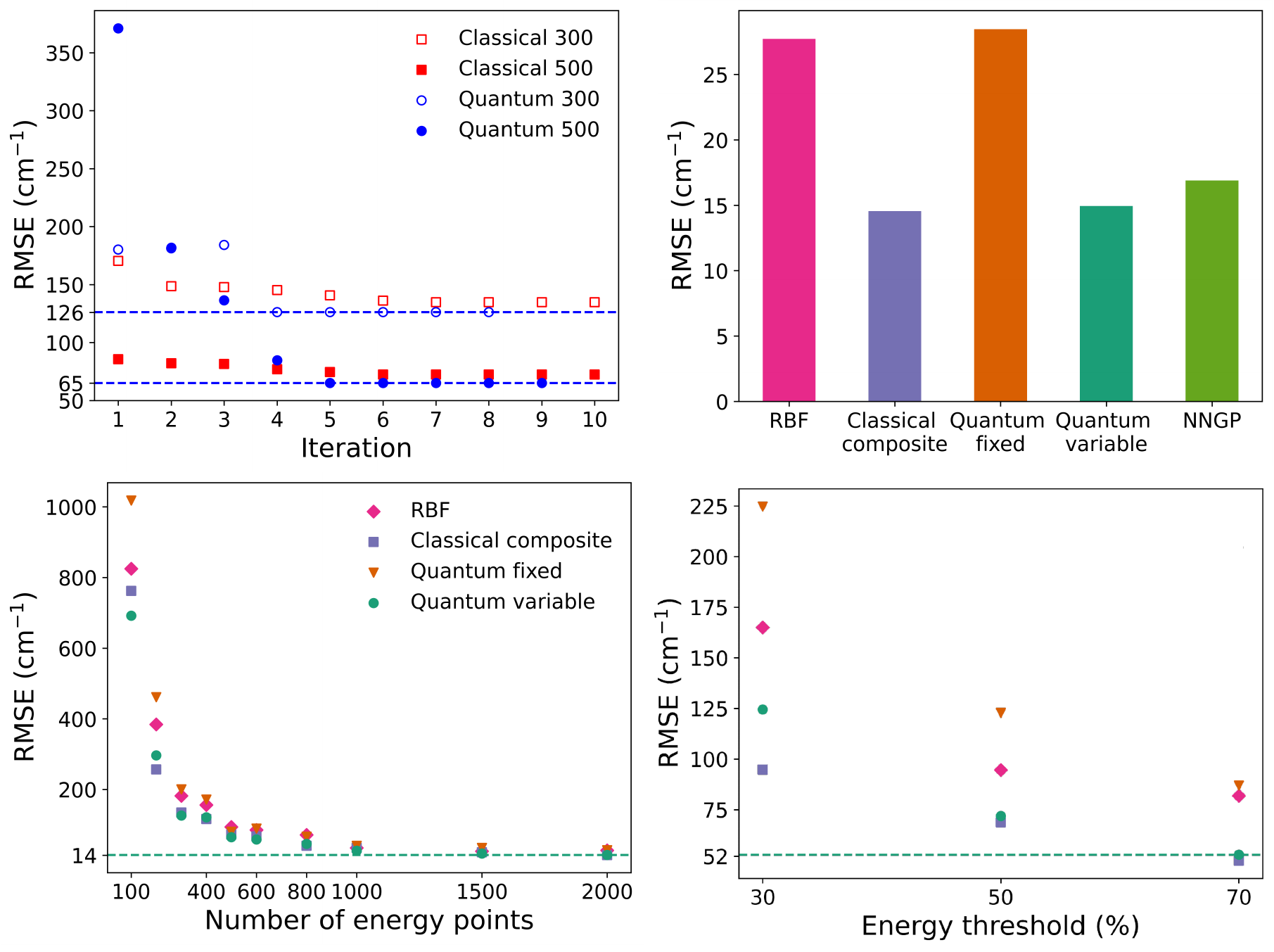}
	\caption{\label{molecule2} Same as Figure \ref{molecule1} but for the molecule H$_2$CO. 
	}
\end{figure*}

\begin{figure*}
	\centering
	\includegraphics[width=\textwidth]{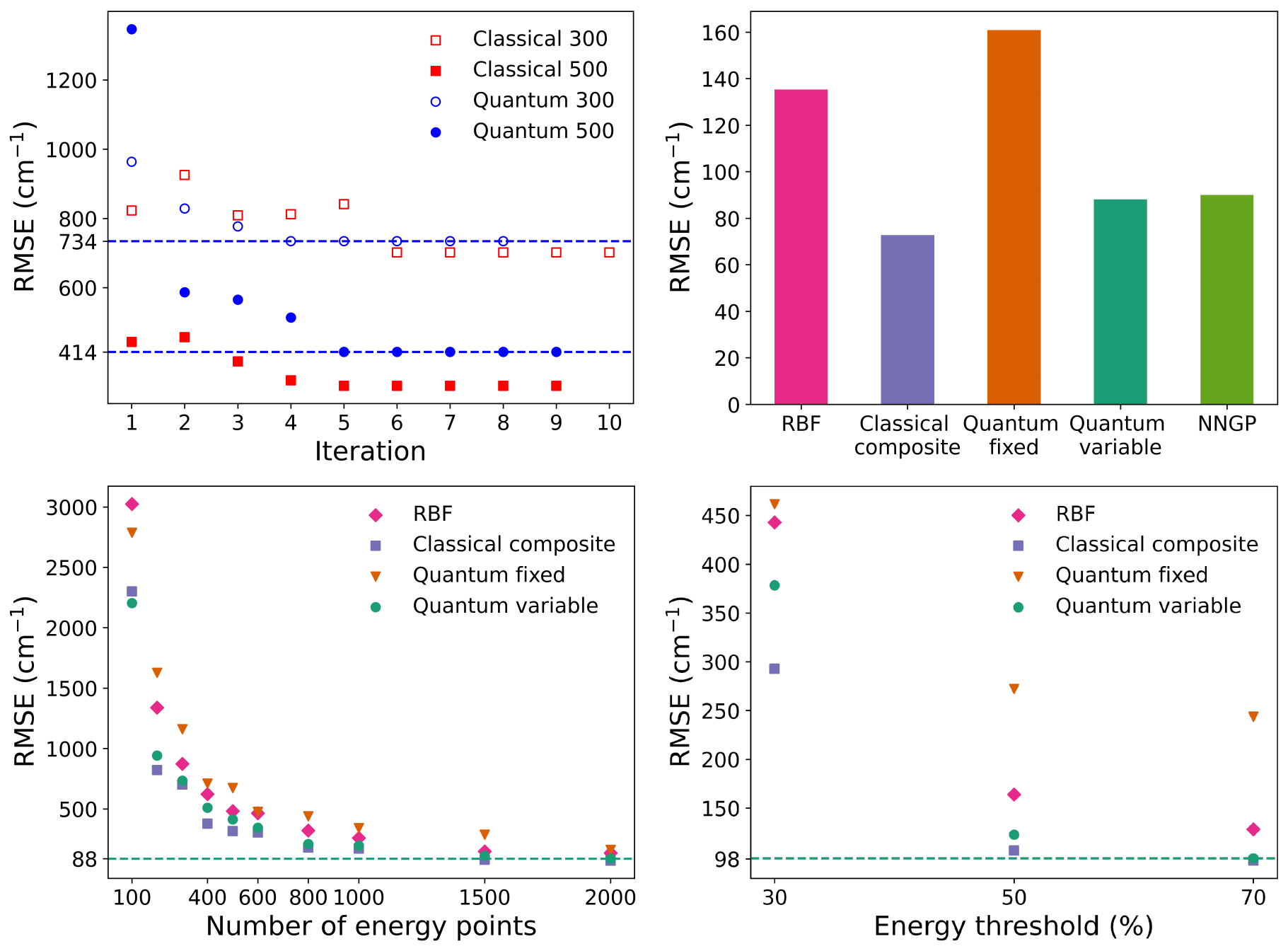}
	\caption{\label{molecule3} Same as Figure \ref{molecule1} but for the molecule HNO$_2$.
	}
\end{figure*}

The compositional search of gate sequences aims to produce quantum kernels aligned with the target functions of GP regression. 
In the limit of $M \rightarrow \infty$ (c.f. Figure \ref{fig:comp_search}), the compositional search is designed to yield an optimal quantum circuit for a given data set.  
In this section, we benchmark the performance of the resulting quantum kernels (obtained with $M = 75$) by comparison of the resulting quantum GP models with GP models employing two kinds of optimized classical kernels: composite kernels with the functional form determined by the algorithm of Duvenaud and coworkers \cite{NIPS2011_4c5bde74,pmlr-v28-duvenaud13,vargas-hernandezExtrapolatingQuantumObservables2018,daiInterpolationExtrapolationGlobal2020} and NNGP kernels. The results are presented in Figures (\ref{molecule1}) - (\ref{molecule2}) for three different regression problems. The figures depict the results for the models of the global PES for H$_3$O$^+$ (Fig.  \ref{molecule1}), H$_2$CO (Fig.  \ref{molecule2}), and HNO$_2$ (Fig.  \ref{molecule3}). 

The upper left panel of each of these figures shows the convergence of the model error with the complexity of the quantum kernels and the classical composite kernels. For quantum kernels, the number of iterations corresponds to the number of layers in the underlying quantum circuit. Each iteration adds one single-layer circuit in the most optimal position as determined by the algorithm. 
For classical composite kernels, the number of iterations corresponds to the depth of the search tree depicted in Figure 1. Each iteration adds one simple kernel function to the complex mathematical form as depicted by Figure \ref{fig:composite_classical}. 
It can be seen that the errors of the classical and quantum models of PES converge to the same value as the number of iterations increases. 
This is illustrated with two training samples (300 and 500 energy data points) used both for the compositional search of the kernels and for training the models for each PES. 
This indicates that quantum kernels can achieve similar expressivity as classical kernels for regression problems.

The upper right panel of Figures (\ref{molecule1}) -- (\ref{molecule3}) compares the errors of the most accurate models of PES for H$_3$O$^+$ (Fig.  \ref{molecule1}), H$_2$CO (Fig.  \ref{molecule2}) and HNO$_2$ (Fig.  \ref{molecule3}) based on a single RBF kernel, classical composite kernels, quantum kernels with the fixed ansatzs from Refs. \onlinecite{daiQuantumGaussianProcess2022,havlicekSupervisedLearningQuantumenhanced2019}, quantum kernels with the adaptive, variable ansatz, and NNGP kernels. Where the structure of the kernels is adaptive, the kernels are built with 2000 energy points. 
The GP models are then built with these kernels and with 2000 energy points in the training set. The results show that RMSE of the quantum modes with the variable, adaptive ansatz yield similar accuracy as the most performant classical kernels for all three cases considered.  

This observation is further illustrated in the lower panels of  Figures (\ref{molecule1}) - (\ref{molecule2}) showing the dependence of RMSE for interpolation (lower left) and extrapolation (lower right) models of PES. The interpolation models are built with energy point samples from the entire energy range of PES. 
To ensure unbiased comparison, these samples are selected to be identical for each kernel, but random for each number of data points considered. The structure of the kernels (for composite classical and quantum variables) is determined with the number of energy points indicated on the $x$-axis randomly sampled for each kernel.  The same energy points are used for determining the kernel hyperparameters and for GP model predictions.

The results shown in the lower right panel of Figures (\ref{molecule1}) -- (\ref{molecule3}) illustrate the error of GP models that extrapolate PES in the energy domain. The 1500 training data samples for these models are randomly chosen from a fixed energy range below the energy threshold specified on the $x$ axis as a percentage of the energy range of the full PES. RMSE is then calculated using all available potential energy points from the energy range above the energy threshold, i.e. the energy range outside of the training data distribution. The results show that GP models with classical composite kernels and quantum kernels based on the variable, adaptive ansatze yield similar extrapolation accuracy. At the same time, the extrapolation performance of these models is much better than the performance of GP models with the RBF kernel or with quantum kernels using the fixed ansatz.

\clearpage
\newpage

\section{Summary}

The main goal of the present work is to present an unbiased comparison of the performance of classical kernels and quantum kernels for regression problems of practical interest. 
To perform such a comparison, it is necessary to construct the most optimal classical kernels and the most optimal quantum kernels for a given data set.  
We seek to determine if quantum kernels can achieve the same expressivity as classical kernels for practical regression problems. 
This is an open question because the previous studies considered either model low-dimensional problems or comparisons between quantum kernels and specific, fixed classical kernels. 

For classical kernels, we adopt two strategies. 
The first strategy increases the complexity of the kernel function incrementally by an iterative approach designed to maximize the Bayesian information criterion. This leads to composite kernel functions designed to maximize the generalization accuracy of a given regression model. The second, independent, strategy employs neural network Gaussian processes, which allows for enhancing the kernel expressivity by increasing the number of NN layers. We note that, while the two approaches yield models with similar accuracy, the composite kernel functions are consistently more accurate for the problems considered here. 
This indicates that the composite kernel functions determined by BIC maximization provide robust benchmarks for quantum kernels. 

For quantum kernels, we develop an algorithm that uses an analog of BIC to optimize the sequence of quantum gates used to estimate quantum fidelities. We show that this algorithm yields quantum circuits producing much better accuracy with fewer quantum gates than a fixed ansatz proposed and used previously. The present algorithm increases the complexity of the quantum circuits incrementally while improving the performance of the resulting kernels.  The algorithm is controlled by hyperparameters that balance the efficiency of the circuit architecture optimization and the size of the explored space of gate permutations.A similar strategy can be used for other applications that require optimization of gate sequences. For example, a closely related algorithm was used in our previous work for state preparation in order to identify efficient quantum circuit representations of ro-vibrational states of polyatomic molecules for computations with variational quantum eigensolvers \cite{asnaashari2023compactquantumcircuitsvariational}. More generally, various algorithms based on incremental, iterative growth of quantum circuits have been employed in variational computations for state preparation \cite{duQuantumCircuitArchitecture2022,9773233,pmlr-v202-wu23v,pmlr-v202-lu23f} and can potentially be exploited for quantum control \cite{GE2022314}. 
The present algorithm can be used for applications in Refs. \onlinecite{duQuantumCircuitArchitecture2022,9773233,pmlr-v202-wu23v,pmlr-v202-lu23f,GE2022314} and the 
algorithms for circuit growth in Refs. \onlinecite{duQuantumCircuitArchitecture2022,9773233,pmlr-v202-wu23v,pmlr-v202-lu23f,GE2022314} can be adapted for quantum regression problems.

 Our results show that the errors of the regression models with classical composite kernels and quantum variable circuits converge to the same value as the complexities of the kernels increase. This indicates that quantum kernels can achieve similar, though not better, expressivity as classical kernels for regression problems. We have considered interpolation models yielding global six-dimensional PES for three polyatomic molecules with training data from the entire energy range and extrapolation models that use {\it ab initio} data from a low-energy part PES to predict PES at higher energies. For extrapolation models, classical composite kernels and quantum variable kernels yield the best models with similar accuracy, significantly exceeding the accuracy of models with RBF kernels or the accuracy of quantum models with a fixed ansatz.  

Finally, our work demonstrates that quantum kernels can be used to build accurate models of global PES for polyatomic molecules. The interpolation RMSE of the 6D PES obtained with a random distribution of 2000 energy points is 16 cm$^{-1}$ for H$_3$O$^+$, 15 cm$^{-1}$ for H$_2$CO and 88 cm$^{-1}$ for 
HNO$_2$. These errors can be further reduced by increasing the number of energy points used for model training. 
This indicates that fitting of PES for molecular systems can be performed on a quantum computer.  
When combined with previously proposed quantum algorithms for quantum chemistry \cite{barkoutsosQuantumAlgorithmAlchemical2021,bauerQuantumAlgorithmsQuantum2020,caoQuantumChemistryAge2019b, delgadoVariationalQuantumAlgorithm2021,hastingsImprovingQuantumAlgorithms2014,lanyonQuantumChemistryQuantum2010,liuQuantumSimulationQuantum2020,maMultiscaleQuantumAlgorithms2023,mottaEmergingQuantumComputing2022,singhBenchmarkingDifferentOptimizers2023,sugisakiQuantumChemistryQuantum2016,yungIntroductionQuantumAlgorithms2014}, and nuclear dynamics \cite{bedard-hearnMeanfieldDynamicsStochastic2005,kassalPolynomialtimeQuantumAlgorithm2008,kovyrshinNonadiabaticNuclearElectron2023,ollitraultNonadiabaticMolecularQuantum2020a,ollitraultMolecularQuantumDynamics2021b,yeter-aydenizPracticalQuantumComputation2020}, this implies that the full stack of {\it ab initio} calculations of molecular properties, including electronic structure, fitting of PES and molecular dynamics, can be implemented on a quantum computer. 
\begin{acknowledgments}
This work was supported by the Natural Sciences and
Engineering Research Council (NSERC) of Canada.
\end{acknowledgments}

\clearpage
\newpage

\bibliography{refs} 

\end{document}